\documentclass[twocolumn,prl,floatfix,superscriptaddress]{revtex4}
\usepackage[T1]{fontenc}
\usepackage{amsmath,amsfonts,amssymb,color,epsfig,graphics,graphicx,latexsym,theorem,url,verbatim}

\usepackage{mathptmx}

\begin{document} \title{Experimental demonstration of wave-particle duality relation based on coherence measure}

\author{Yuan Yuan}
\affiliation{CAS Key Laboratory of Quantum Information, University of Science and Technology of China, Hefei, 230026, People's Republic of China}
\affiliation{Synergetic Innovation Center of Quantum Information and Quantum Physics, University of Science and Technology of China, Hefei, Anhui 230026, People's Republic of China}
\author{Zhibo Hou}
\affiliation{CAS Key Laboratory of Quantum Information, University of Science and Technology of China, Hefei, 230026, People's Republic of China}
\affiliation{Synergetic Innovation Center of Quantum Information and Quantum Physics, University of Science and Technology of China, Hefei, Anhui 230026, People's Republic of China}
\author{Yuan-Yuan Zhao}
\affiliation{CAS Key Laboratory of Quantum Information, University of Science and Technology of China, Hefei, 230026, People's Republic of China}
\affiliation{Synergetic Innovation Center of Quantum Information and Quantum Physics, University of Science and Technology of China, Hefei, Anhui 230026, People's Republic of China}
\author{Han-Sen Zhong}
\affiliation{CAS Key Laboratory of Quantum Information, University of Science and Technology of China, Hefei, 230026, People's Republic of China}
\affiliation{Synergetic Innovation Center of Quantum Information and Quantum Physics, University of Science and Technology of China, Hefei, Anhui 230026, People's Republic of China}
\author{Guo-Yong Xiang}
\email{gyxiang@ustc.edu.cn}
\affiliation{CAS Key Laboratory of Quantum Information, University of Science and Technology of China, Hefei, 230026, People's Republic of China}
\affiliation{Synergetic Innovation Center of Quantum Information and Quantum Physics, University of Science and Technology of China, Hefei, Anhui 230026, People's Republic of China}
\author{Chuan-Feng Li}
\affiliation{CAS Key Laboratory of Quantum Information, University of Science and Technology of China, Hefei, 230026, People's Republic of China}
\affiliation{Synergetic Innovation Center of Quantum Information and Quantum Physics, University of Science and Technology of China, Hefei, Anhui 230026, People's Republic of China}
\author{Guang-Can Guo}
\affiliation{CAS Key Laboratory of Quantum Information, University of Science and Technology of China, Hefei, 230026, People's Republic of China}
\affiliation{Synergetic Innovation Center of Quantum Information and Quantum Physics, University of Science and Technology of China, Hefei, Anhui 230026, People's Republic of China}
\

\begin{abstract}
Wave-particle duality is a typical example of Bohr's complementarity principle that plays a significant role in quantum mechanics. Previous studies used the visibility of an interference pattern to quantify the wave property and used path information to quantify the particle property. However, coherence is the core and basis of the interference phenomenon. If we could use coherence to characterize the wave property, the understanding of wave-particle duality would be strengthened. A recent theoretical work [Phys. Rev. Lett. 116, 160406 (2016)] found two relations between quantum coherence and path information. Here, we demonstrate the new measure of wave-particle duality based on two kinds of coherence measures quantitatively for the first time. The wave property, quantified by the coherence in the $l_{1}$-norm measure and the relative entropy measure, can be obtained via tomography of the target state, which is encoded in the path degree of freedom of the photons. The particle property, quantified by the path information, can be obtained via the discrimination of detector states, which is encoded in the polarization degree of freedom of the photons. Our work may deepen people's understanding of coherence and provide a new perspective regarding wave-particle duality.
\end{abstract}

\maketitle

\section{Introduction}
Bohr's complementarity principle is at the heart of quantum mechanics. The core of this principle is that an object has multiple properties and that these properties cannot, in theory, be measured simultaneously \cite{bohr1,bohr2}. A typical example is wave-particle duality, which is one of famous yet intriguing features of quantum mechanics. A particle that goes through an interferometer can exhibit either wave or particle properties. The particle properties are characterized by information regarding which path a particle takes, while the wave properties are characterized by the visibility of the interference pattern. If we have complete information regarding which path a particle takes, the interference visibility is zero, and if the interference visibility is at its maximum value of one, we have no information regarding which path a particle takes. There are, of course, intermediate situations in which some path information leads to a reduction in the interference visibility, which have been studied experimentally \cite{WP1,WP2,WP3,WP4}. The intermediate case was first investigated by Wooters and Zurek in 1978 \cite{WZ}. In 1988, Greenberger and Yasin found an inequality that expresses the tradeoff between interference visibility and path information \cite{zhongzi},
\begin{equation}\label{v}
V^{2}+D^{2}\leq1,
\end{equation}
where $V$ is the visibility of the interference pattern and $D$ is a measure of the path information, i.e., the path distinguishability or the which-path information. Englert, in a seminal paper, added detectors to the scenario, and obtained a relation of the form given in Eq. (\ref{v}) \cite{Englert}. Different states of the detectors correspond to different paths, and Englert's measure of path information, $D$, is based on one's ability to distinguish the different detector states. The inequality given in Eq. (\ref{v}) has been verified experimentally using cold neutral atoms \cite{atom}, nuclear magnetic resonance (NMR) \cite{NMS1,NMS2}, a faint laser \cite{laser}, and single photons in a delayed-choice scheme \cite{photon}. In addition, many theoretical and experimental studies of the path-visibility relation in interferometers with more than two paths were proposed in \cite{Jaeger,Mpath1,Mpath2,Mpath5,Mpath6}.

Recently, two measures of quantum coherence have been proposed \cite{coherence}. These measures resulted from the resource theory of quantum coherence, which has stimulated several further studies \cite{property1,property2,distillation1,distillation2,frozen1,frozen2}. The first application of the recently defined coherence measures to wave-particle duality relations was done by Bera \emph{et al.}, who used the $l_1$ coherence measure to quantify the wave nature of a particle \cite{firstPRA}. The path information was characterized by an upper bound of the probability of successfully discriminating the detector states by means of unambiguous state discrimination. Subsequently, two new duality relations were found \cite{PRL}. The wave nature was characterized by coherence in the $l_{1}$ measure and entropy measure. The path information was characterized by the maximum probability of successfully discriminating the detector states by the method of minimum-error state discrimination.

In previous works, the visibility of the interference pattern has been widely used to quantify the wave nature of an object. However, coherence is at the heart of interference phenomenon \cite{coherence}. As an alternative candidate, the quantified coherence can be used as a generalization of the interference visibility to characterize the wave property of a particle inside a multipath interferometer, which would provide a new perspective to quantify wave property and deepen the understanding of the quantum coherence and wave-particle duality. Until now, these new measure of wave-particle duality relations have not yet been demonstrated experimentally. To fill this gap, we experimentally demonstrate the wave-particle duality relation based on coherence measures. Here, the wave property and particle property of the photons are studied using a Mach-Zehnder interferometer. The detector states are encoded in the polarization degree of freedom of the photons, and the state of the photons (target state) is encoded in path degree of freedom of the photons. The particle property of the photons can be characterized by path information using minimum-error state discrimination and by the mutual information between detector states and the outcome of the measurement performed on them; the wave property of the photons can be characterized by the coherence in the $l_{1}$ measure and relative entropy measure using the tomography of the target state.

\begin{figure*}[tbph]
\begin{center}
\includegraphics [width=18cm,height=5cm]{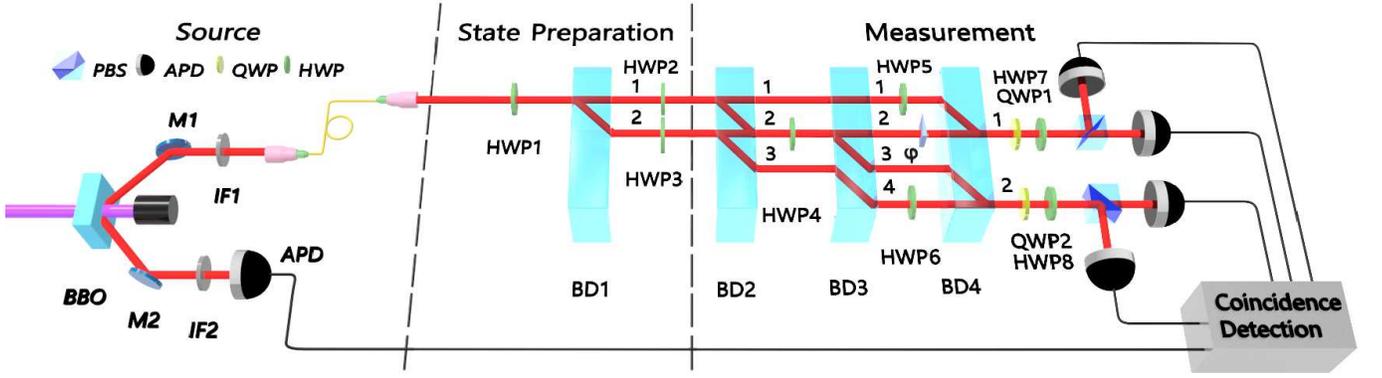}
\end{center}
\caption{Experimental setup. In the source module, the single photon source is generated by the spontaneous parametric down-conversion(SPDC) progress with a type-II beamlike phase-matching beta-barium-borate(BBO). In the state preparation module, the angle of HWP1 is $22.5^{\circ}$; thus, the photons pass through path-1 and path-2 equiprobably. The detector states can be changed by rotating HWP2 and HWP3. In the measurement module, we measure coherence and path information independently. The angle of HWP4 is $0^{\circ}$ or $45^{\circ}$, depending on the observation of coherence via tomography of the target state or observation of the path information via discrimination of the detector states. HWP5 and HWP6 are both rotated by $45^{\circ}$ in an effort to rotate V polarized to H polarized and rotate H polarized to V polarized, respectively. The phase $\varphi=180^{\circ}$ is added only in the wave property measurement. Quarter-wave plate (QWP)1, QWP2, HWP7, HWP8 and PBS are used to perform tomography and optimal minimum-error state discrimination. Output photons are detected using avalanche photo-diode(APD).}
\label{fig:map}
\end{figure*}

\section{Relation between coherence in entropy measure and path information}
\subsection{Theory}
Considering a photon entering a Mach-Zehnder interferometer, the state of the photon is in the superposition state $|\psi\rangle=\frac{1}{\sqrt{2}}(|1\rangle+|2\rangle)$, where the orthonormal basis states $|1\rangle$ and $|2\rangle$ correspond to path-1 and path-2 in the interferometer. To obtain the path information of the photons, a detector is introduced in each path and the path information is encoded in the detector states of $|\eta_{1}\rangle$ and $|\eta_{2}\rangle$. Whether the photons pass through path-1 or path-2 is determined by the results of discriminating the detector states \cite{Englert}. If the detector state is found to be $|\eta_{1}\rangle$, the photon is considered to pass through path-1; otherwise, it passes through path-2. After the photons interact with the detector system, the state of the entire system is expressed as $|\Psi\rangle=\frac{1}{\sqrt{2}}(|1\rangle|\eta_{1}\rangle+|2\rangle|\eta_{2}\rangle)$. Here, $|\eta_{1}\rangle$ and $|\eta_{2}\rangle$ are encoded in the polarization degree of freedom of the photons and can be defined as
\begin{equation}\label{detector state}
\begin{aligned}
&|\eta_{1}\rangle=\cos\theta|H\rangle+\sin\theta|V\rangle,
\\&|\eta_{2}\rangle=\cos\theta|H\rangle-\sin\theta|V\rangle,
\end{aligned}
\end{equation}
where $|H\rangle$ is the horizontal polarization and $|V\rangle$ is the vertical polarization. Tracing out the detector, we find that the density matrix of photons (target state) is given by
\begin{equation}\label{photonp}
\rho=Tr_{det}(|\Psi\rangle\langle\Psi|)=\begin{pmatrix} \frac{1}{2} & \frac{1}{2}\cos2\theta \\\frac{1}{2}\cos2\theta &  \frac{1}{2} \end{pmatrix}.
\end{equation}
The relative entropy measure of the coherence for a density matrix $\rho$ is given by $C(\rho)=S(\rho_{diag})-S(\rho)$, where $\rho_{diag}$ is a diagonal density matrix of $\rho$ and the von Neumann entropy is $S(\rho)=-Tr(\rho\log_{2}\rho)$. Thus, the theoretical value of coherence of the target state $C(\rho)$ is expressed as
\begin{equation}\label{C}
C(\rho)=1+(\cos^{2}\theta\log_{2}\cos^{2}\theta+\sin^{2}\theta\log_{2}\sin^{2}\theta).
\end{equation}
We use coherence $C(\rho)$ to characterize the wave property of the photons, which is one quantity that needs to be measured in the experiment.

To obtain the path information, we adopt the widely-used method of minimum-error state discrimination to probe the detector states \cite{book}. We will quantify the path information by the mutual information $H(M:D)$ between the detector states labeling the paths and the results of probing them. The mutual information $H(M:D)$ is defined as $H(M:D)=H(D)+H(M)-H({p_{ij}})$, where $D$ is a variable corresponding to the detector states $|\eta_{i}\rangle$ and $M$ is a variable corresponding to the measurement result of state discrimination. $H({p_{i}})=-\sum_{i=1}^{2}p_{i}\log p_{i}$ is the Shannon entropy. The detector state can be obtained by tracing out the photons, given by $\rho_{det}=\frac{1}{2}(|\eta_{1}\rangle\langle\eta_{1}|+|\eta_{2}\rangle\langle\eta_{2}|)$. So the detector state $\rho_{i}=|\eta_{i}\rangle\langle\eta_{i}|$ appears with a probability of $p_{i}=1/2$; thus, its information content is $H(D)=H({p_{i}})=1$. To obtain maximum mutual information $H(M:D)$, we will probe detector states using the optimal positive operator valued measure(POVM) $\Pi_{i}$ to discriminate them and thereby determine the path that the photons pass through. For the detector states $|\eta_{1}\rangle$ and $|\eta_{2}\rangle$, the optimal measurement is a projective measurement $\Pi_{1}=|\phi_{1}\rangle\langle\phi_{1}|$ and $\Pi_{2}=|\phi_{2}\rangle\langle\phi_{2}|$ \cite{discrimination}, where
\begin{equation}\label{M}
\begin{aligned}
&|\phi_{1}\rangle=\frac{1}{\sqrt{2}}(|H\rangle+|V\rangle),
\\&|\phi_{2}\rangle=\frac{1}{\sqrt{2}}(|H\rangle-|V\rangle).
\end{aligned}
\end{equation}
The probability of measurement results $M=i$ is $p(M=i)=Tr(\Pi_{i}\rho_{det})$, so the information content of $H(M)$ is given by $H(M)=H({q_{i}})=1$, where $q_{i}\equiv p(M=i)$. The joint distribution of the two variables $D$ and $M$ is written as $p_{ij}\equiv p(M=i, D=j)=Tr(\Pi_{i}\rho_{j})p_{j}$. Thus, the theoretical value of the maximum mutual information $H(M:D)$ is expressed as
\begin{equation}\label{H}
H(M:D)=2+2p_{11}\log_{2}p_{11}+2p_{12}\log_{2}p_{12},
\end{equation}
where $p_{11}=p_{22}=\frac{1}{4}+\frac{1}{4}\sin2\theta$ and $p_{12}=p_{21}=\frac{1}{4}-\frac{1}{4}\sin2\theta$. If the two variables are perfectly correlated, the mutual information $H(M:D)$ is $H(D)$, while if they are uncorrelated, the mutual information is equal to zero. We use the mutual information $H(M:D)$ to characterize the particle property of the photons, which is the other quantity that needs to be measured in the experiment. Theoretically, it is proved that $C(\rho)$ and $H(M:D)$ satisfy the following duality relation \cite{PRL},
\begin{equation}\label{second relation}
C(\rho)+H(M:D)\leq H({p_{i}}).
\end{equation}
In our model, the bound of the inequality is $H({p_{i}})=1$. This entropic version of the coherence-path information duality relation in the case of two-path will be demonstrated in the following experiment.

\subsection{Experimental setup}
The whole experimental setup is shown in Fig. 1. It consists of three modules: single-photon source module (see Methods for details), state preparation module and measurement module. In the state preparation module, the beam displacer (BD) causes the vertical polarized component to be transmitted directly and the horizontal polarized component to undergo a 4-mm lateral displacement. A photon in the horizontal polarized state passes through a half-wave plate (HWP) HWP1 with rotation angle $22.5^{\circ}$, the state of $\frac{1}{\sqrt{2}}(|H\rangle+|V\rangle)$ is generated. We label the paths that emerge after each BD, from the top to the bottom, as 1, 2, \emph{ect.}, as shown in Fig. 1. After the photon  passes through BD1, the state of the whole system can be written as $\frac{1}{\sqrt{2}}(|1\rangle|V\rangle+|2\rangle|H\rangle)$. The state of the photons (target state) is encoded in the path degree of freedom of the photons, and the states of the detector are encoded in the polarization degree of freedom of the photons. HWP2 in path-1 rotates the polarization of the photons from $|V\rangle$ to $|\eta_{1}\rangle$, and HWP3 in path-2 rotates the polarization of the photons from $|H\rangle$ to $|\eta_{2}\rangle$. Hence, after passing through HWP2 and HWP3, the photons interact with the detector system, and the state of the entire system can be prepared as $|\Psi\rangle=\frac{1}{\sqrt{2}}[|1\rangle\otimes(\cos\theta|H\rangle+\sin\theta|V\rangle)+|2\rangle\otimes(\cos\theta|H\rangle-\sin\theta|V\rangle)]$, with an arbitrary $\theta$ ranging from $0^{\circ}$ to $90^{\circ}$ because of the rotation of HWP2 and HWP3.

In the measurement module, by the interferences between different BDs, we can obtain desired states which are used to perform wave property and particle property measurements. By setting angle of HWP4 to $0^{\circ}$ or $45^{\circ}$, we can measure coherence via tomography of the target state or measure the path information via discrimination of the detector states. To observe the wave property of the photons, the tomography of the target state should be performed. It is difficult to perform tomography in the path basis state directly; thus, we try to transform it to polarization basis state. The angle of HWP4 should be set to $0^{\circ}$. After BD2, the state of the entire system becomes $\frac{1}{\sqrt{2}}[\sin\theta|V\rangle|1\rangle+(\cos\theta|H\rangle-\sin\theta|V\rangle)|2\rangle+\cos\theta|H\rangle|3\rangle]$. After BD3, the state of the entire system becomes $\frac{1}{\sqrt{2}}[\sin\theta|V\rangle|1\rangle+\sin\theta|V\rangle|2\rangle+\cos\theta|H\rangle|3\rangle+\cos\theta|H\rangle|4\rangle]$. HWP5 and HWP6 both are rotated by $45^{\circ}$ in an attempt to make the photons in path-1 and path-2 overlap at BD4; the same is true for path-3 and path-4. There are two interferences between BD1 and BD4. Adding a phase $\varphi=180^{\circ}$ in the path-2, the final state of entire system becomes $\frac{1}{\sqrt{2}}[\sin\theta(|H\rangle-|V\rangle)|1\rangle+\cos\theta(|H\rangle+|V\rangle)|2\rangle]$ after BD4. Actually, the ensemble of devices from BD2 to BD4 can be expressed as $U=[\sigma_{x}H]_{target}\bigotimes [H]_{detector}$, where $H$ is Hadamard gate and given by $H=\frac{1}{\sqrt{2}}[1,1;1,-1]$. Finally, we perform tomography in the polarization basis with QWP1, QWP2, HWP7, HWP8 and PBS, and obtain two density matrices. The density matrix of the target state can be expressed as the sum of the weights of two density matrices, which is used to calculate the coherence of the photons.

To observe the particle property of the photons, the detector states should be discriminated, as shown in Eq. (\ref{detector state}). The angle of HWP4 should be set to $45^{\circ}$, and thereby the ensemble of devices from BD2 to BD4 and HWPs with angle $45^{\circ}$ in path-1 and path-2 of BD4 (not shown in Fig. 1) can be expressed as $U=I_{4\times4}$, where $I$ is identity matrix. Therefore, the detector states, are transformed by BD2-BD4, will be the same as the initially prepared detector states ($|\eta_{1}\rangle$, $|\eta_{2}\rangle$). Finally, we use optimal measurement to probe detector states in the polarization basis state, and obtain the joint distribution of the two variables $D$ and $M$ and their respective probability distributions. The optimal measurement is constructed in Eq. (\ref{M}), and realized by HWP7, HWP8 and PBS.

We obtain an interference visibility of 0.996 for each interference. Because HWP2 and HWP3 are inside the two interferometers between BD1 and BD4, the interference visibility will be decreased when HWP2 and HWP3 are rotated every time. In order to keep high interference visibility, we need to adjust interferometer once HWP2 and HWP3 are rotated. Because of the small separations between the neighboring paths, it is a challenge to inset HWP4 (4mm$\times$4mm) with setting desired angles in the middle path without influencing the photons in the neighboring paths.

\begin{figure}[tbph]
\begin{center}
\includegraphics [width=9cm,height=9cm]{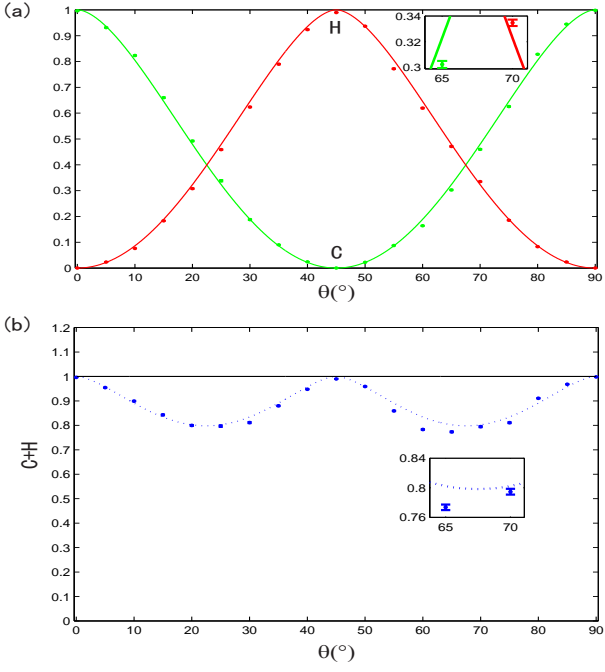}
\end{center}
\caption{Experimental results regarding the relation between coherence in the relative entropy measure and the path information. Figure 2(a) shows the coherence $C$ (green) and the mutual information $H$ (red) between the detector states labeling the paths and the results of probing them as a function of the detector states $\theta$. The solid lines are the theoretical expectations Eqs. (\ref{C}) and (\ref{H}). Figure 2(b) shows the sum of coherence $C$ and the mutual information $H$; the blue dashed line denotes the theoretical values of $C+H$. The black solid line represents the upper bound of inequality Eq. (\ref{second relation}). The error is too small to identify using this coordinate system; thus, we provide partially enlarged drawings using a magnifying power of 5x.}
\label{fig:result2}
\end{figure}

\subsection{Experimental result}
Based on the measurement described above, we can obtain the coherence in the relative entropy measure and the maximum mutual information of the photons for different detector states, as depicted in Fig. 2. Figure 2(a) shows that the experimental results of $H$ and $C$ are in good agreement with the theoretical values. In the case of $\theta=0^{\circ}$, a maximum coherence of $C_{max}=0.9963\pm0.0007$ is observed, while no mutual information is obtained. In the case of $\theta=45^{\circ}$, a maximum mutual information of $H_{max}=0.9897\pm0.0008$ is obtained, while no coherence is observed. In the case of intermediate values $\theta$, the increase in mutual information $H$ leads to a decrease in coherence $C$, and each quantity varies from 0 to 1. This is consistent with the notion of Bohr's complementarity principle. Different from previous wave-particle duality relation Eq. (\ref{v}), figure 2(b) shows that the sum of $C$ and $H$ attains to the bound of inequality Eq. (\ref{second relation}) only in the case of $\theta=0^{\circ}$ and $\theta=45^{\circ}$.

\section{Relation between coherence in $l_{1}$ measure and path information}
We also demonstrate the relation between the coherence in $l_{1}$ measure and the path information \cite{PRL},
\begin{equation}\label{first relation}
(P_{s}-\frac{1}{N})^{2} + X^{2} \leq (1-\frac{1}{N})^{2},
\end{equation}
where $P_{s}$ represents the maximum probability of successfully identifying the detector states using the optimal POVM measurement and characterizes the particle property of the photons,
and $X$ represents the coherence in $l_{1}$ measure and characterizes the wave property of the photons. The coherence in the $l_1$ measure of the state is given by $X=(1/N)C_{l_{1}}(\rho)$, where $C_{l_{1}}(\rho)=\sum_{i,j=1,i\neq j}^N |\rho_{ij}|$ and $N$ is the number of interferometer paths. In our case, $N$ should be 2. According to Eq. (\ref{photonp}), the theoretical value of coherence $X$ of the photons is
\begin{equation}\label{X}
X=\frac{1}{2}\cos2\theta.
\end{equation}
The maximum average probability of successfully identifying the detector states is given  by
\begin{equation}\label{ps}
P_{s}=\sum_{i=1}^{2}\frac{1}{2}\langle\eta_{i}|\Pi_{i}|\eta_{i}\rangle=\frac{1}{2}+\frac{1}{2}\sin2\theta.
\end{equation}
If we have no prior information about the path that the photons pass through, the probability that we will guess correctly is $1/N$. Thus, $P_{s}-1/N$ can be understood as the measure of how much better we can do by using detectors than by only guessing. The experimental results regarding the observed coherence in $l_{1}$ measure and the path information are depicted in Fig. 3.

\begin{figure}[tbph]
\begin{center}
\includegraphics [width=9cm,height=9cm]{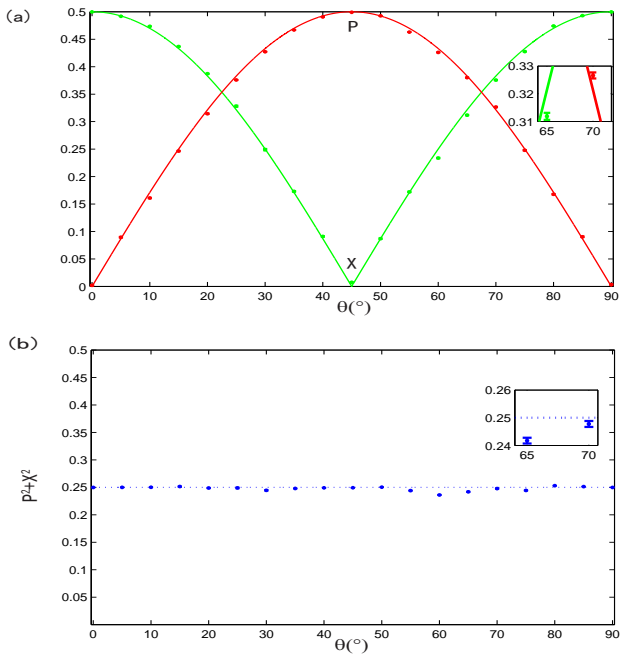}
\end{center}
\caption{Experimental results for the relation between coherence in the $l_{1}$ measure and path information. Figure 3(a) shows coherence $X$ (green) and path distinguishability $P$ (red) as a function of the detector states $\theta$. Here, we define $P=P_{s}-1/2$. The solid lines are the theoretical expectations Eqs. (\ref{X}) and (\ref{ps}). Figure 3(b) shows the sum of the squares of coherence and the squares of path distinguishability. The blue dashed line denotes the theoretical values. We provide partially enlarged drawings using a magnifying power of 5x.}
\label{fig:result1}
\end{figure}

From Fig. 3(a), we can see that no which-way information is obtained, while a maximum coherence of $X_{max}=0.4997\pm0.0001$ is observed at $\theta=0^{\circ}$. In the case of $\theta=45^{\circ}$, the full which-way information of $P_{max}=0.4991\pm0.0001$ is obtained, while no coherence is observed. In the case of $\theta$ at intermediate values, incomplete which-way information is obtained, and partial coherence is observed. Each quantity varies from 0 to 0.5, and the increase in path information $P$ leads to a decrease in coherence $X$. Figure 3(b) plots the sum of $X^{2}$ and $P^{2}$, which is invariant with respect to $\theta$ and thereby demonstrates the relation Eq. (\ref{first relation}) is actually an equality for $N=2$. Meanwhile the experimental results imply that the new duality relation Eq. (\ref{first relation}) can return to Eq. (\ref{v}) in the case of two-path interferometer. Experimental results coincide, but the physical meanings of visibility and coherence are inherently different. Coherence plays an indispensable role in quantum resource theory, and is the core and basis of the interference phenomenon. Moreover, compared with the previous experimental works of demonstration of the wave-particle duality relation \cite{atom,NMS1,NMS2,laser,photon}, we experimentally obtain wave property of the photons by measuring the coherence in stead of interference visibility.

\section{Conclusion}
To summarize, previous research tried to use a number of different quantities as a generalization of the interference visibility, but coherence is the core of the interference phenomenon; thus, as an alternative candidate, the quantified coherence could be used as a generalization of the visibility to describe the wave property. In this paper, we perform the first quantitative test of recently proposed of wave-particle duality relations based on coherence measures using a Mach-Zehnder interferometer. In the experiment, we obtain the coherence and path information of a photon for different detector states, which are applied to the interferometer. Our experimental results agree well the theoretical predictions. The bound of the inequality in $l_{1}$ measure can be obtained using any kind of detector state, while the bound of the inequality in entropy measure can only be obtained using the orthogonal detector states. In any case, the two new measure duality relations are both tight and sustained the test in a two-path interferometer. Our experimental results suggest that the quantified coherence in the $l_{1}$ norm and the relative entropy measures are strong candidates for the generalization of the interference visibility to describe the wave property in a multipath interferometer. Our work not only could deepen people's understanding of quantum coherence but also provides a new perspective regarding the wave-particle duality relation.

\section{Methods}
\subsection{The details for the single-photon source module of experimental setup}
In the single-photon source module, a 80-mW cw laser with a 404-nm wavelength (linewidth=5MHz) pumps a type-II beamlike phase-matching beta-barium-borate (BBO, 6.0$\times$6.0$\times$2.0 mm$^{3}$, $\theta=40.98^{\circ}$) crystal to produce the degenerate photon pairs. After being redirected by mirrors M1 and M2 and passing through the interference filters (IF, $\bigtriangleup\lambda$=3 nm, $\lambda$=808 nm), the photon pairs generated in the spontaneous parametric down-conversion (SPDC) process are coupled into single-mode fibers separately.
The total coincidence counts are approximately $5\times10^{3}$ every second.

\subsection{Error analysis}
The error of the experimental results is estimated via a Monte Carlo simulation based on the photon number detection with the Poisson distribution. The measurement time is 10s per data point, so the error bar of the experimental results is relatively small. The maximum error bar and minimum error bar are $\pm0.0042$ and $\pm0.0001$, respectively. Hence, the errors in our experiment mainly come from the inaccuracy of angles of the wave plates and the imperfect interference visibility of the interferometer.

\section{Acknowledgments}
We thank Prof. Mark Hillery and Mr. Meng-Jun Hu for the helpful discussion regarding the organization of the manuscript. This work is supported by the National Natural Science Foundation of China (NSFC) (11574291, 11774334) and National Key R \& D Program (2016YFA0301700).

\end{document}